\documentclass[aps,prb,twocolumn,showpacs,showkeys,preprintnumbers,groupedaddress,amsmath,amssymb,]{revtex4-1}

\usepackage{graphicx}% Include figure files
\usepackage{dcolumn}% Align table columns on decimal point
\usepackage{bm}% bold math
\usepackage[dvips]{color}

\begin{document}


\title{Granular hopping conduction in (Ag,Mo)$_x$(SnO$_2$)$_{1-x}$ films in the dielectric regime}

\author{Ya-Nan Wu}
\author{Yan-Fang Wei}
\author{Zhi-Qing Li}
\email[Electronic address: ]{zhiqingli@tju.edu.cn}

\affiliation{Tianjin Key Laboratory of Low Dimensional Materials Physics and Preparing Technology, Department
of Physics, Tianjin University, Tianjin 300072, China}

\author{Juhn-Jong Lin}
\email[Electronic address: ]{jjlin@mail.nctu.edu.tw}

\affiliation{Institute of Physics and Department of Electrophysics, National Chiao Tung University, Hsinchu
30010, Taiwan}

%\date{\today}

\begin{abstract}

We have studied the temperature dependence of conductivity, $\sigma(T)$, in two series of
Ag$_{x}$(SnO$_2$)$_{1-x}$ and Mo$_{x}$(SnO$_2$)$_{1-x}$ granular films lying below the percolation threshold,
where $x$ is the metal volume fraction. The metal grains in the former series have approximately spherical
shape, whereas in the latter series contain numerous small substructures. In both series of films, we have
observed the $\sigma \propto \exp[-(T_0/T)^m]$ temperature dependence, with $m\,\simeq\,1/2$, over the wide
$T$ range from 2 to 80 (100) K, where $T_0$ is a characteristic temperature. Our $\sigma$($T$) results are
explained in term of the Abeles-Sheng model which considers the structural effect of metal grains. The
extracted values of the optimal separation between grains ($s_{m}$) and the charging energy ($E_{c}$) slightly
increase with decreasing $x$. The variation of the parameter $C(x)$, defined in the Abeles-Sheng model, is in
consistency with a simplified expression which depends only on $x$ and the dielectric constant of the
insulating matrix. As the temperature is increased to above $\sim$\,100 K, a crossover to the thermally
fluctuation-induced tunneling conduction processes is observed.

\end{abstract}

\keywords{granular hopping conduction, dielectric regime, metal-insulator composites, percolation}

\maketitle

\section{Introduction}

Granular metals, $M_xI_{1-x}$, are composite mixtures composed of nanometer-sized metal particles embedded in
a dielectric matrix, where $M$ denotes a metal, $I$ denotes an insulator, and $x$ denotes the metal volume
fraction. Due to the specific nanoscale structure, granular metals reveal novel physical properties that are
not found in homogeneous systems.\cite{Rev. Mod. Phys.79469} For instance, the electron-electron interaction
effect on the transport properties in the presence of granularity has been demonstrated to be significantly
different from those in homogeneous disordered metals.\cite{PRB84 052202,PRB91 104201} A $\ln (T)$ temperature
dependence of conductivity and Hall coefficient was observed to be independent of granular array
dimensionality. The transport properties of granular systems in the dielectric regime ($x<x_c$, the
percolation threshold) are expected to be distinctive as well. Very often, the following temperature
dependence of electrical conductivity, $\sigma$, is observed over an extended range of
$T$:\cite{PRL35247WAlO,PRB361962,PRB13931, NeugebauerWebb,PRB2612,PSheng5712, TChuiAl,SolidStateCommun,
CJAdkins7143, CJAdkins1253,PhilosMagB855,EatinWohlmanVRH, PRB035411,Linnanotech045711,IBalberg035318}
\begin{equation}\label{Eq.(sigmma)}
\sigma(T)=\sigma_{0} \exp\left[ -\left( T_0/T\right)^{1/2} \right] \,\,,
\end{equation}
where $\sigma_0$ and $T_0$ are sample dependent parameters. On the other hand, the $T$ dependence as described
by the form of Eq. (1) has also been widely observed in amorphous and doped crystalline semiconductors, and
satisfactorily explained by the Efros-Shklovskii variable-range-hopping (VRH) conduction processes in the
presence of a Coulomb gap.\cite{ESVRH1,ESVRH2,crossover1,crossover2,crossover3}

As early as in the 1970s, Abeles, Sheng and coworkers have explored this kind of stretched-exponential
$\sigma(T)$ behavior, Eq. (1), by considering the structural effects of constituting metal
grains.\cite{PSheng3144,PSheng24407,PSheng2583} They have noted that charge carriers are generated by thermal
activation and transferred by tunneling between neighboring grains. The tunneling takes place along certain
optimal percolation paths. The selection of optimal paths is based on an assumption that the ratio
\emph{s}/\emph{a} is a function of $x$ alone, where \emph{s} is the separation between neighboring metal
grains, and \emph{a} is the metal grain size. This theory has been successfully used to explain the results in
a number of granular systems that have a distribution of $a$.\cite{PRL35247WAlO,PRB361962,Linnanotech045711}
However, it has also been argued that the Abeles-Sheng model is inappropriate to account for the observations
in other granular systems.\cite{TChuiAl,periodic1,periodicgold} Multiple theoretical models aiming at
clarifying the robust $\sigma \propto \exp[-(T_0/T)^{1/2}]$ behavior have been put forward over years, but a
full theory has yet to be achieved.\cite{SolidStateCommun, CJAdkins7143,CJAdkins1253,PhilosMagB855} Recently,
a correlated hopping mechanism based on multiple co-tunneling processes has been formulated, where the
electrostatic disorder induced by charged traps in the insulating matrix is predicted to cause the form of Eq.
(1).\cite{PRB125121} The validity of this co-tunneling model awaits experimental tests.

To clarify this long-standing problem, we study the $\sigma(T)$ behaviors of two series of
Ag$_x$(SnO$_2$)$_{1-x}$ and Mo$_x$(SnO$_2$)$_{1-x}$ granular films in this work. The first series of films
contains approximately spherical metal grains with a grain size distribution smaller than $\approx$\,40\%. The
second series of films has small substructures in the Mo grains. The electronic transport properties for these
films in the metallic regime ($x\,>\,x_c$) were reported previously.\cite{doublepercolation,GHEMo} We point
out that the percolation threshold is $x^{c}_{Ag}\simeq0.50$ in the Ag$_x$(SnO$_2$)$_{1-x}$ films and
$x^{c}_{Mo}\simeq0.32$ in the Mo$_x$(SnO$_2$)$_{1-x}$ films.

\section{Experimental method}

The cosputtering fabrication method for our samples was described previously.\cite{doublepercolation,GHEMo}
The thickness of the films was determined by a surface profiler (Dektak, 6 M). The Ag$_x$(SnO$_2$)$_{1-x}$
films were $\approx$\,500 nm thick, and Mo$_x$(SnO$_2$)$_{1-x}$ films were $\approx$\,350 nm thick. The Ag
(Mo) volume fraction \emph{x} in each sample was obtained from the energy-dispersive x-ray spectroscopy
analysis. The microstructure of the films was studied by the transmission electron microscopy (TEM, Tecnai G2
F20). The electrical measurements were performed on a physical property measurement system (PPMS-6000, Quantum
Design), employing the four-probe technique.

\begin{figure}[htp]
\begin{center}
\includegraphics[scale=1.05]{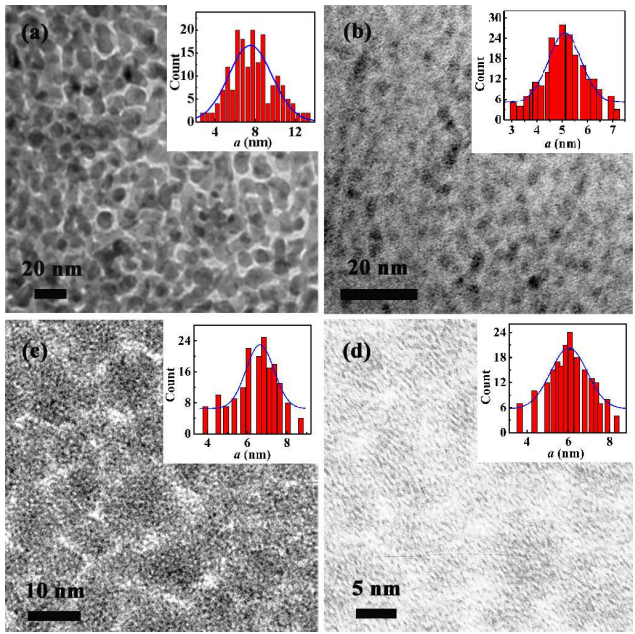}
\caption{TEM images for Ag$_x$(SnO$_2$)$_{1-x}$ films with (a) $x\,\simeq$ 0.62 (taken from Ref.
\onlinecite{PRB91 104201}), and (b) $\simeq$\,0.26 (taken from Ref. \onlinecite{doublepercolation}); and for
Mo$_x$(SnO$_2$)$_{1-x}$ films with (c) $x\simeq$\,0.36, and (d) $\simeq$\,0.29 [(c) and (d) are taken from Ref.
\onlinecite{GHEMo}]. The inset in each panel shows the corresponding grain size distribution histogram. The
percolation thresholds are $x^{c}_{Ag}\simeq0.50$ and $x^{c}_{Mo}\simeq0.32$ in Ag$_x$(SnO$_2$)$_{1-x}$ and
Mo$_x$(SnO$_2$)$_{1-x}$ films, respectively.}
\label{TEM}
\end{center}
\end{figure}

\section{Results and discussion}

Figure \ref{TEM} shows representative TEM images for two Ag$_x$(SnO$_2$)$_{1-x}$ films and two
Mo$_x$(SnO$_2$)$_{1-x}$ films, with $x$ as indicated in the caption to Fig. \ref{TEM}. The inset in each panel
shows the corresponding grain size distribution histogram. We note that the two series of films have distinct
metal grain structures. The Ag$_x$(SnO$_2$)$_{1-x}$ films reveal typical granular
characteristics.\cite{doublepercolation} The Ag particles are approximately spherical in shape and embedded in
the amorphous SnO$_2$ matrix, with well-defined metal grain boundaries. On the other hand, in the
Mo$_x$(SnO$_2$)$_{1-x}$ films, there are no regularly shaped Mo grain boundaries. Both Mo and SnO$_2$
particles reveal small substructures penetrating into each other.\cite{GHEMo} Selected-area diffraction
patterns indicate that the Ag grains are crystalline,\cite{doublepercolation,PRB91 104201} whereas the Mo
grains are amorphous.\cite{GHEMo} The mean grain size of Ag and Mo grains in Ag$_x$(SnO$_2$)$_{1-x}$ and
Mo$_x$(SnO$_2$)$_{1-x}$ films with $x<x_c$ are $\approx$\,5$\pm$2 nm and $\approx$\,6$\pm$4 nm, respectively,
see Figs. \ref{TEM}(b) and \ref{TEM}(d). It is worth noting that these mean grain sizes are slightly smaller
than those in the films with $x>x_c$, see Figs. \ref{TEM}(a) and \ref{TEM}(c).

\begin{figure}[htp]
\begin{center}
\includegraphics[scale=1.05]{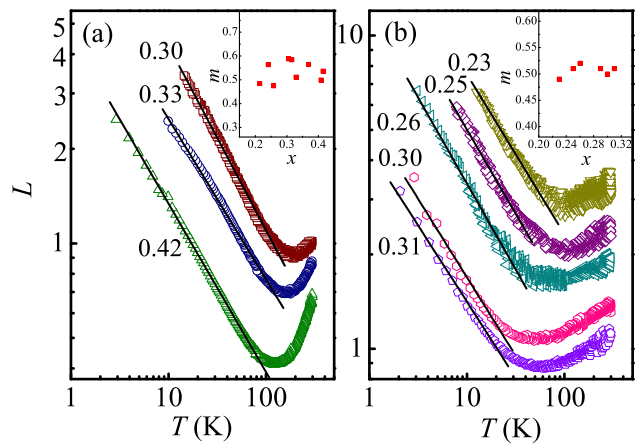}
\caption{Parameter $L$ versus $T$ in log-log scales for several (a) Ag-SnO$_2$
films, and (b) Mo-SnO$_2$ films, with the metal volume fraction $x$ as indicated. The solid straight lines are
linear fits. Insets: The fitted slope $m$ as a function of $x$ for each series of films.}
\label{LT}
\end{center}
\end{figure}

Assume that the temperature $T$ dependence of conductivity can be written in the form $\sigma(T)$ =
$\sigma_{0}\exp[-(T_{0}/T)^{m}]$. We define a parameter $L(T) \equiv
\partial\log_{10}(\sigma)/\partial\log_{10}(T) = -m(T_{0}/T)^{m}$, and then write
$\log_{10}L(T)$ = $\log_{10}(-mT_{0}^{m}) -m\log_{10}T$.\cite{wt} That is, there exists a linear expression
for $\log_{10}L$ versus $\log_{10}T$, with $m$ as the slope. In Fig. \ref{LT}, we plot the parameter $L$
calculated from our measured $\sigma(T)$ in double logarithmic scales for both series of films, as indicated.
Obviously, $\log_{10}L$ varies linearly with $\log_{10}T$ over a wide range of $T$ for all the representative
films shown in Figs. \ref{LT}(a) and \ref{LT}(b). We thus obtain the slopes $m$ for each series of films and
plot them in the corresponding insets. It is clearly seen that $m$ takes a value of $\simeq$\,0.5, which in
turn strongly suggests the validity of the $\sigma \propto \exp[-(T_0/T)^{1/2}]$ temperature law. Thus, it is
well justified to perform least-squares fits of our $\sigma(T)$ data to Eq. (1). We have then extracted the
values of the adjusting parameters $\sigma_0$ and $T_0$ for each film, as listed in Table I.

\begin{table}
\caption{\label{tableT0}Relevant parameters for Ag$_x$(SnO$_2$)$_{1-x}$ and Mo$_x$(SnO$_2$)$_{1-x}$ films.
$\sigma_{0}$ and $T_{0}$ are defined in Eq. (1), $T_{1}$ and $T_{0F}$ are defined in Eq. (5), and $\phi_{0}$
($w$) is the barrier height (width) computed from Eqs. (6) and (7) by setting the effective barrier area
$A\approx$\,3 nm$^{2}$.}
\begin{ruledtabular}
\begin{tabular}{ccccccccc}%\hline \hline
    & $x$ & $\sigma(300\texttt{ K})$ & $\sigma_{0}$ & $T_{0}$ & $T_{1}$ & $T_{0F}$ & $\phi_{0}$ & $w$  \\
       &  & (S/m) & (S/m) & (K) & (K) &(K) & (meV) & ({\AA}) \\  \hline
Ag & 0.21 & 132 & 415 & 728 & 2018 & 341 & 145 & 19.3     \\
series& 0.24 & 126 & 442 & 679 & 1825 & 333 & 135 & 18.5  \\
   & 0.28 & 288 & 1598 & 727 & 1538 & 331 & 118 & 16.8    \\
   & 0.30 & 848 & 3912 & 613 & 794 & 207 & 84.1 & 16.4    \\
   & 0.31 & 779 & 3085 & 445 & 793 & 236 & 79.8 & 14.8    \\
   & 0.33 & 1506 & 3248 & 245 & 816 & 267 & 77.7 & 13.6    \\
   & 0.37 & 1760 & 3942 & 199 & 720 & 276 & 69.4 & 12.3    \\
   & 0.41 & 2973 & 4144 & 77.4 & 652 & 297 & 62.3 & 10.9    \\
   & 0.42 & 3133 & 4459 & 71.3 & 610 & 294 & 59.2 & 10.6    \\   \hline
Mo & 0.23 & 74.8 & 183 & 2056  & 2706 & 214 & 222 & 33.4    \\
series& 0.25 & 170 & 259 & 942 & 2205 & 239 & 180 & 27.1    \\
   & 0.26 & 662 & 661 & 470 & 1443 & 198 & 138 & 24.4       \\
   & 0.29 & 574 & 442 & 465  & 1352 & 181 & 136 & 25.2      \\
   & 0.30 & 1337 & 597 & 107 & 1241 & 237 & 114 & 19.3      \\
   & 0.31 & 2528 & 1657 & 88.2 & 827 & 203 & 87.6 & 17.1    \\
\end{tabular}
\end{ruledtabular}
\end{table}

\subsection{Comparison with the Abeles-Sheng model}

In the Abeles-Sheng model, the $T$ dependence of conductivity originates from the optimization of the product
of mobility and density of charges over all percolation paths. Abeles and Sheng have defined the parameter
$C\equiv\chi sE_{c}$, where $\chi=\sqrt{2m\phi/\hbar^{2}}$ is a constant related to the effective barrier
height $\phi$, and $E_c$ is the charging energy required to generate a pair of positively and negatively
charged grains. Abeles and Sheng have obtained the form of Eq. (1) with the relation $C=k_BT_0/4$, where $k_B$
is the Boltzmann constant. Thus, the value of $C$ for a given sample is experimentally determined from the
extracted value of $T_0$.

\begin{figure}[htp]
\begin{center}
\includegraphics[scale=1.08]{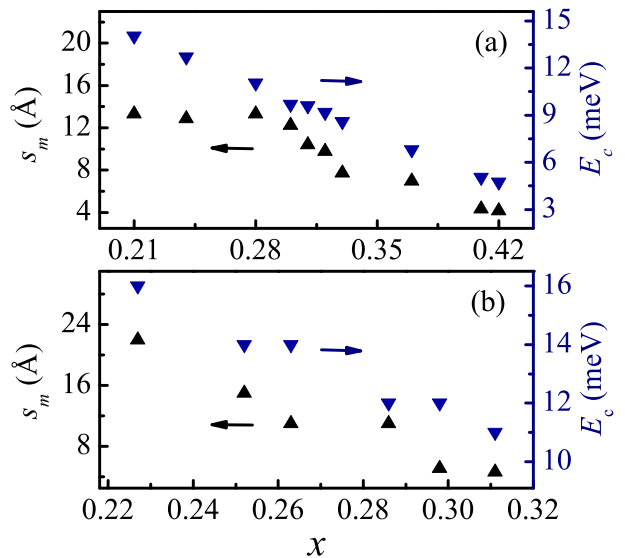}
\caption{Optimal separation between metal grains $s_m$ and charging energy $E_c$ computed from the
Abeles-Sheng model for (a) Ag$_x$(SnO$_2$)$_{1-x}$ films, and (b) Mo$_x$(SnO$_2$)$_{1-x}$ films. The values of
$s_m$ are for $T$ = 10 K.}
\label{smEc}
\end{center}
\end{figure}

Abeles and Sheng have assumed that the ratio $s/a$ is a function of $x$ alone. By further assuming that the
metal grains are spherical and packed in a simple cubic lattice, they have obtained the expression
\begin{equation}\label{Eq.(sax)}
s/a=\left(\pi/6x\right)^{1/3}-1 \,\,.
\end{equation}
They have calculated the optimal separation between neighboring grains in the percolating paths, denoted by
$s_m$, and obtained $s_m=\sqrt{C/k_BT}/2\chi$. To compare the experimental data with their theoretical
predictions, we take $\phi\approx$\,0.1 eV in our films. This value is just the difference between the work
function of Ag (Mo) $\simeq$\,4.6 eV and the electron affinity of SnO$_2$ $\simeq$\,4.5
eV.\cite{PSheng2834,workfunctionAg,workfunctionMo,affinitySnO2} Then, we have the estimate of $\chi$\,=\,0.16
{\AA}$^{-1}$. The values of $s_m$ for our films can then readily be extracted from the known values of $C$ and
$\chi$. Figures \ref{smEc}(a) and \ref{smEc}(b) show the variation of $s_m$ (at a representative $T$\,=\,10 K)
with $x$ for Ag$_x$(SnO$_2$)$_{1-x}$ and Mo$_x$(SnO$_2$)$_{1-x}$ films, respectively. We see that, in both
series of films, $s_m$ slightly increases from 0.4 to 2.2 nm with decreasing $x$. This variation of $s_m$ is
in line with that previously found in other composite
materials.\cite{Linnanotech045711,PSheng24407,PSheng2834}

The charging energy is defined by $E_{c}=e^{2}/2\pi\epsilon_{0}\kappa a$, where $\kappa=\epsilon(1+a/2s)$ is
the effective dielectric constant in the granular composite, and $\epsilon_{0}$ ($\epsilon$) is the
permittivity of vacuum (insulating medium). Using $\epsilon$ = 12 for the amorphous SnO$_2$
matrix,\cite{1dielectricSnO2,2dielectricSnO2} we have obtained the variation of $E_c$ with $x$, as also shown
in Fig. \ref{smEc}. The typical value of $E_c$ for those films lying just below the percolation threshold is
on the order of a few meV.

\begin{figure}[htp]
\begin{center}
\includegraphics[scale=1]{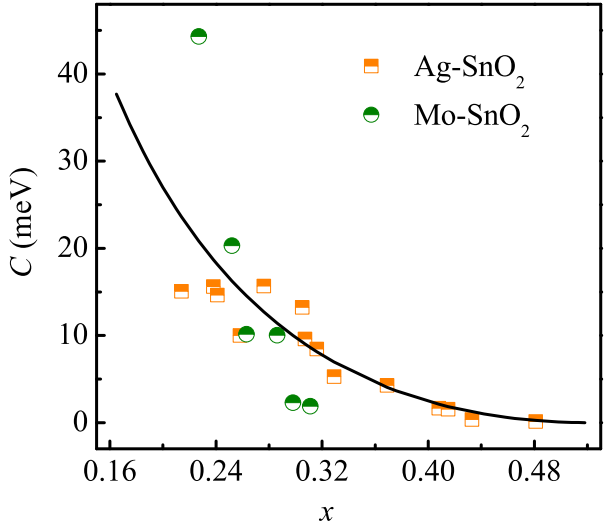}
\caption{Variation of parameter $C$ with $x$ for Ag$_x$(SnO$_2$)$_{1-x}$ and Mo$_x$(SnO$_2$)$_{1-x}$ films, as
indicated. The symbols are the experimental data, and the solid curve is the prediction of Eq. (3) with
$\eta$\,=\,0.17 eV.}
\label{cx}
\end{center}
\end{figure}

The structural and composition dependent parameter $C$ defined in the Abeles-Sheng model can be expressed in
an explicit form:\cite{PSheng24407}
\begin{equation}\label{Eq.(Cx)}
C=\eta\,\frac{\left(s/a\right)^{2}}{\left(1/2\right)+\left(s/a\right)}\,\,,
\end{equation}
where
\begin{equation}\label{Eq.(eta)}
\eta=\chi e^{2}/2\pi\epsilon_{0}\epsilon \,\,.
\end{equation}
This parameter can provide a self-consistency check for the validity of the model, since it depends only on
$x$ and the dielectric constant of the insulating matrix. Figure \ref{cx} shows the variation of $C$ with $x$
for the Ag$_x$(SnO$_2$)$_{1-x}$ and Mo$_x$(SnO$_2$)$_{1-x}$ films, as indicated. The solid curve is the
least-squares fit to Eq. (3) with the single adjusting parameter $\eta$ = 0.17 eV. Inspection of Fig. \ref{cx}
indicates that the results for the Ag$_x$(SnO$_2$)$_{1-x}$ films can be well described by Eq. (3), whereas
there is some discrepancy for the case of the Mo$_x$(SnO$_2$)$_{1-x}$ films. The discrepancy is not
unexpected. It may originate from the irregular shape of Mo grains and a relatively large distribution in
grain size $a$.

\subsection{Fluctuation-induced tunneling conduction above $\sim$\,100 K}

As the temperature is increased to above $\sim$\,100 K, our measured $\sigma(T)$ behavior crosses over to the
fluctuation-induced tunneling (FIT) conduction processes. The FIT mechanism has originally been formulated for
disordered systems characterized by large conducting regions separated by thin insulating barriers. In this
model, the thermally induced voltage fluctuations have modulating effects on potential barriers, which play an
important role in enhancing the electron tunneling probability and thereby lead to a characteristic $T$
dependence of conductivity\cite{FIT1,FIT2,FIT3}
\begin{equation}\label{Eq.(FIT)}
\sigma(T)=\sigma_{0F}\exp \left[ -\frac{T_{1}}{T+T_{0F}} \right] \,\,,
\end{equation}
where $\sigma_{0F}$ is a weakly $T$ dependent parameter. The characteristic temperatures $T_{1}$ and $T_{0F}$
are defined by\cite{T1T0,ZnOnanowire}
\begin{equation}\label{Eq.(T1)}
T_{1}=\frac{8\epsilon_{0}\epsilon A\phi_{0}^{2}}{k_{B}e^{2}w} \,\,,
\end{equation}
and
\begin{equation}\label{Eq.(T0F)}
T_{0F}=\frac{16\epsilon_{0}\epsilon\hbar A\phi_{0}^{3/2}}{\pi(2m_{e})^{1/2}k_{B}e^{2}w^{2}} \,\,,
\end{equation}
where $A$ is the effective tunneling barrier area, $m_{e}$ is the charge carrier mass, $\hbar$ is the Planck
constant divided by 2$\pi$, and $\phi_{0}$ ($w$) is the insulating barrier height (width).

\begin{figure}[htp]
\begin{center}
\includegraphics[scale=1.08]{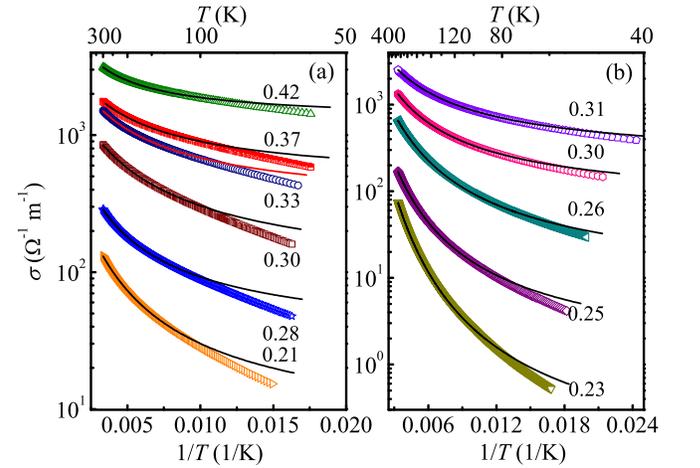}
\caption{Variation of conductivity with inverse temperature for several (a) Ag$_x$(SnO$_2$)$_{1-x}$ films,
and (b) Mo$_x$(SnO$_2$)$_{1-x}$ films. The solid curves are least-squares fits to Eq. (5). The numbers
indicate the metal volume fraction $x$.}
\label{FIT}
\end{center}
\end{figure}

Figures \ref{FIT}(a) and \ref{FIT}(b) show the conductivity as a function of $1/T$ for Ag$_x$(SnO$_2$)$_{1-x}$
and Mo$_x$(SnO$_2$)$_{1-x}$ films above 40 K, respectively. The solid curves are least-squares fits to Eq.
(5). Clearly, the experimental $\sigma(T)$ data for both series of films can be well described by Eq. (5) from
300 K down to $\sim$\,100 K. In the FIT model, the size of conducting grains should be large enough such that
$E_c < k_{B}T$. Thus, the value $T_{min} = E_{c}/k_{B}$ indicates a low-bound temperature for the FIT
mechanism to be applicable. We evaluate $T_{min} \approx$\,110 K for the Ag$_x$(SnO$_2$)$_{1-x}$ films and
$\approx$\,150 K for the Mo$_x$(SnO$_2$)$_{1-x}$ films. A good agreement between this estimate and the
experiment is reached for the Ag-based films, because the metal grains in this series of films have spherical
shape with a fairly well defined grain size. The agreement for the Mo-based films is not as good, but still
acceptable.

According to the FIT model, most of the electron tunneling occurs within the small surface areas of metallic
grains. Hence, the effective barrier area $A$ should be given by the size of the closest approach between
grains. Assuming that the value of $A$ to be on the order of a tenth of the maximal cross section of a
spherical grain, we take $A \sim$\,3 nm$^{2}$ to compute the values of $\phi_{0}$ and $w$ in Eqs. (6) and (7).
The calculated values are listed in Table I. We see that the values of $\phi_{0}$ and $w$ decrease with
increasing $x$. Our $\phi_{0}$ values are somewhat lower than those extracted for C-PVC composites,\cite{FIT2}
and RuO$_{2}$ and IrO$_{2}$ nanowires,\cite{nanowireLin} but compatible to those in ZnO
nanowires,\cite{ZnOnanowire} micrometer-sized Al/AlO$_x$/Y tunnel junctions,\cite{Lai} and ITO
films.\cite{ITOfilm} This smallness of the extracted $\phi_{0}$ value has recently been theoretically
reconsidered by Xie and Sheng.\cite{Xie} On the other hand, the extracted $w$ values are in good consistency
with the $s_{m}$ values extracted from the Abeles-Sheng model discussed in the above subsection, and similar
to those reported for the above-mentioned materials.

\section{Conclusion}

We have studied the electrical transport properties of two series of Ag$_x$(SnO$_2$)$_{1-x}$ and
Mo$_x$(SnO$_2$)$_{1-x}$ granular films in the dielectric regime. In all films, the $\sigma \propto
\exp[-(T_0/T)^{1/2}]$ temperature dependence has been observed over a wide temperature range from 2 to
$\sim$\,100 K. We explain our results within the framework of the Abeles-Sheng model that considers structural
effect of metal grains. The extracted values of the relevant parameters are in satisfactory agreement with the
theory. As the temperature is increased to above about 100 K, the conductivity is governed by the thermally
fluctuation-induced tunneling conduction processes.

\acknowledgments

This work was supported by the National Natural Science Foundation of China (NSFC) through Grant No. 11174216
and the Research Fund for the Doctoral Program of Higher Education through Grant No. 20120032110065 (Z.Q.L.),
and by the Taiwan Ministry of Science and Technology through Grant No. NSC 103-2112-M-009- 017-MY3 and the MOE
ATU Plan (J.J.L.).

%\bibliography{WuYanan}

\end{document}